\documentclass[%
 reprint,
 superscriptaddress,
 amsmath,amssymb,
 aps,
 prl,
]{revtex4-2}

\usepackage{overpic}
\usepackage{graphicx}
\usepackage{dcolumn}
\usepackage{tikz}
\usepackage{bm}

\begin{document}

\title{Adiabatic sheath model for beam-driven blowout plasma channels}

\author{Yulong Liu}
\affiliation{Institute of High Energy Physics, Chinese Academy of Sciences, Beijing 100049, China}
\affiliation{University of Chinese Academy of Sciences, Beijing 100049, China}

\author{Ming Zeng}
\email[Corresponding author: ]{zengming@ihep.ac.cn}
\affiliation{Institute of High Energy Physics, Chinese Academy of Sciences, Beijing 100049, China}
\affiliation{University of Chinese Academy of Sciences, Beijing 100049, China}

\author{Lars Reichwein}
\affiliation{Institut f\"ur Theoretische Physik I, Heinrich-Heine-Universit\"at D\"usseldorf, D-40225 D\"usseldorf, Germany}
\affiliation{Peter Gr\"unberg Institut (PGI-6), Forschungszentrum J\"ulich, 52425 J\"ulich, Germany}

\author{Alexander Pukhov}
\affiliation{Institut f\"ur Theoretische Physik I, Heinrich-Heine-Universit\"at D\"usseldorf, D-40225 D\"usseldorf, Germany}

\date{\today}

\begin{abstract}
In plasma wakefield accelerators, the structure of the blowout sheath is vital for the blowout radius and the electromagnetic field distribution inside the blowout. Previous theories assume artificial distribution functions for the sheath, which are either inaccurate or require prior knowledge of parameters. In this study, we develop an adiabatic sheath model based on force balancing, which leads to a self-consistent form of the sheath distribution. This model gives a better estimate of the blowout channel balancing radius than previous models.
\end{abstract}

\maketitle

Beam-driven plasma wakefield accelerators (PWFAs) have the capacity to accelerate particles to tens of GeV energies in meter-scale structures, making them competitive candidates for future high-energy accelerators that are used for next-generation particle colliders~\cite{PChenPRL1985,IBlumenfeldNature2007,CAdolphsenarxiv2022,BFosterNJP2023,JGaoRDTM2024}. In a PWFA, when a high current, relativistic electron beam, called the drive beam, interacts with an underdense plasma, plasma electrons are completely radially expelled while the ions remain immobile due to their much larger mass, forming an electron-free ion channel along the propagation axis. The physics of plasma response in this regime is strongly nonlinear and the regime has been referred to as the blowout or bubble regime. In this regime, the acceleration (or deceleration) field is uniform and the focusing field is linear in regard to the radial offset from the axis~\cite{TKatsouleasPRA1986,JJSuIEEE1987,JBRosenzweigPRA1991,NBarovPRE1994,WLuPOP2006}.

On the boundary of the bubble, the expelled electrons form a narrow sheath, where the electron density and current profiles steepen with a large value and decay in a small thickness~\cite{JBRosenzweigPRL1987,JBRosenzweigPRAB2004,IKostyukovPOP2004}. The sheath structure depends on the properties of the drive beam and is strongly correlated with the creation of the plasma blowout, the structure of wakefield inside the blowout, and the processes of particle injection into the wakefield~\cite{MTzoufrasPPL2008,IKostyukovPRL2009,AAGolovanovPOP2016}.

Existing theories have used rectangular or exponential distributions to simplify the sheath shape~\cite{WLuPRL2006, SAYiPOP2013,JThomasPOP2016,AAGolovanovQE2016,AAGolovanovPOP2017,AAGolovanovPPCF2021}. \textcite{TNDalichaouchPOP2021} proposed a multisheath model, which employs two sheath layers to describe the wakefield more accurately. However, these models require the prior knowledge of the sheath thickness, which cannot be obtained self-consistently. Recently, \textcite{AAGolovanovPRL2023} have developed a blowout theory from the energy conservation point of view. They have assumed a $\delta$-function distribution for the blowout sheath, and found the evolution function for the blowout channel radius $r_{\delta}$ as
\begin{equation}
    A \left(r_\delta\right) r_\delta \frac{d^2r_{\delta}}{d \xi^2} + B\left(r_\delta\right) r^2_\delta \left(\frac{dr_{\delta}}{d\xi} \right)^2 + C r^2_\delta = \Lambda, \label{eq:r_delta}
\end{equation}
where $A = 1 + r_{\delta}^2/4$, $B = 1/2 + 1/r_{\delta}^2$ and $C = 1/4$. $\xi=ct-z$ is the longitudinal co-moving coordinate, $c$ is the speed of light in vacuum, $t$ is time, and $z$ is the longitudinal coordinate. The drive term of the equation can be written as $\Lambda = 2I/I_A$, where $I$ is the instant current of the beam, $I_A= 4 \pi \epsilon_0 m_e c^3/e \approx 17\ \rm kA$ is the Alfv\'en current, $\epsilon_0$ is the vacuum permittivity, $m_e$ is the electron rest mass, and $e$ is the elementary charge~\cite{HDWhittumPRA1991, CHuangPRL2007, DLOMartinezPOP2015}. If derivatives of $r_{\delta}$ are assumed to be zero, one can obtain the balancing radius of the channel based on the $\delta$-sheath theory as
\begin{equation}
    r_{\delta0} = 2\sqrt{\Lambda}. \label{eq:r_delta0}
\end{equation}
By contrast, the charge neutralization radius, 
\begin{equation}
    r_n = \sqrt{2\Lambda}, \label{eq:r_n}
\end{equation}
has been widely accepted as the balancing radius from the electrostatic point of view~\cite{GeraciPOP2000, WLuPOP2006, CHuangPRL2007}.

In this work, we develop a self-consistent model for the channel sheath using the adiabatic assumption. The sheath equations are examined in two cases with the help of particle-in-cell (PIC) simulations. We demonstrate that our model provides a more accurate description of blowout channel balancing, as its predicted channel radius falls between the electrostatic neutralization radius and that predicted by the $\delta$-sheath theory.

Throughout the paper, we adopt the plasma normalization units, in which charge is normalized to $e$, mass to $m_e$, velocity to $c$, density to plasma density $n_p$, time to $\omega_p^{-1}$, length to $c/\omega_p$, electric field to $m_e c \omega_p/e$, magnetic field to $m_e \omega_p/e$, electrostatic potential to $m_e c^2/e$ and vector potential to $m_e c/e$, where $\omega_p=\sqrt{e^2 n_p/\epsilon_0 m_e}$ is the plasma frequency.

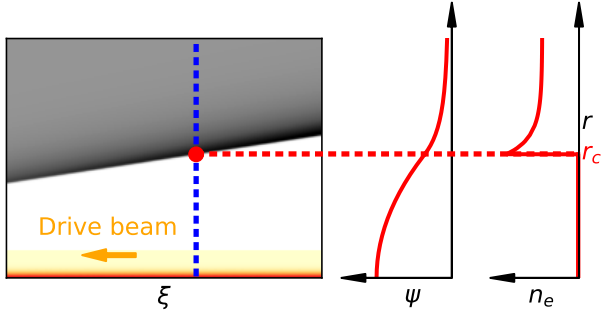
\begin{figure}
    \centering
    \begin{overpic}[width=0.48\textwidth]{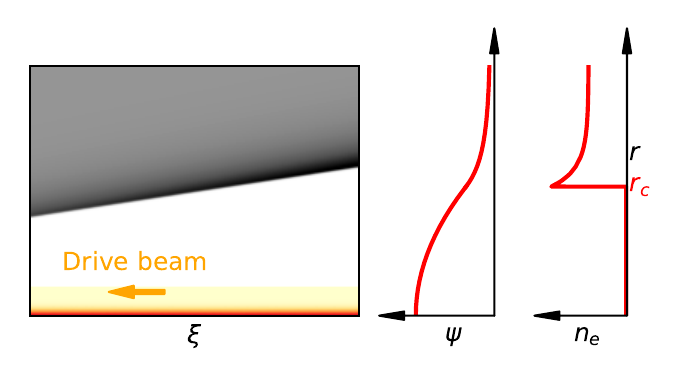}
    \put(33.3,27.55){\begin{tikzpicture}
        \draw[densely dashed, red, line width=2pt] (0,0) -- (5.1,0);
    \end{tikzpicture}}
    \put(32.5,8.6){\begin{tikzpicture}
        \draw[densely dashed, blue, line width=2pt] (0,0) -- (0,3.16);
        \fill[red] (0.00,1.625) circle (3pt);
    \end{tikzpicture}}
    \end{overpic}
    \caption{\label{fig:plamsa_density_and_psi} Illustration of the adiabatic sheath model in this work. The electron beam (orange color, beam head is not shown), with a linear density $\Lambda\left(\xi\right)$, is moving to the left along the $\xi$ axis and drives the plasma wakefield in the blowout regime. The plasma electrons (gray region) are completely evacuated in the channel with the radius $r_c\left(\xi\right)$, leaving an immobile ion column (white region). As $\Lambda$ varies slowly with $\xi$ in our assumption, the transverse electromagnetic balance is achieved everywhere. The right two subplots show the transverse distribution of the pseudo-potential $\psi$ and the electron density $n_e$ at a certain $\xi$ denoted by the blue dashed line.}
\end{figure}

We consider a highly relativistic drive electron beam with density $n_b \left(\xi, r\right)$ that expels the plasma electrons to form an ion channel with a radius $r_c\left(\xi\right)$, where $r$ is the radial coordinate, and cylindrical symmetry is assumed. We assume that the drive beam is narrower than the channel, so that there is no drive charge outside the channel, or $n_b=0$ for $r>r_c$. Under these assumptions, the drive term can also be written as $\Lambda\left(\xi\right) = \int_0^\infty n_b r dr$, which has the physical meaning of the (normalized) linear density of the drive beam.

Furthermore, we assume the radius of the ion channel changes adiabatically, which means $\left(dr_c/r_c\right) /d\xi\ll 1$, $\left(d\Lambda/\Lambda\right) /d\xi\ll 1$, and $\partial_\xi\ll\partial_r$ for all field variables, as illustrated in Fig.~\ref{fig:plamsa_density_and_psi}. Under the cylindrical symmetry and the quasi-static approximation~\cite{DHWhittumPOP1997, PMoraPOP1997}, the pseudo-potential of the wakefield obeys the following Poisson-like equation~\cite{HDWhittumPRL1991, WLuPOP2006}
\begin{equation}
    -\frac{1}{r}\frac{\partial}{\partial r}\left(r\frac{\partial}{\partial r}\psi\right) = S, \label{eq:psi_S}
\end{equation}
where $\psi=\varphi - A_z$ is the pseudo-potential, $\varphi$ is the electrostatic potential, $A_z$ is the longitudinal component of the vector potential,
\begin{equation}
    S = \rho - J_z = 1 - n_e\left(1-v_z\right) \label{eq:S}
\end{equation}
is the source term, $\rho$ is the charge density, $J_z$ is the longitudinal component of the current density, $n_e$ is the plasma electron density, and $v_z$ is the longitudinal velocity of the plasma electrons. Because $n_e=0$ for $r<r_c$, we can write the form of the pseudo-potential inside the blowout
\begin{equation}
    \left.\psi\right|_{r<r_c} = \psi_c + \frac{r_c^2}{4} - \frac{r^2}{4},
\end{equation}
where
\begin{equation}
    \psi_c = \left.\psi\right|_{r=r_c}>0
\end{equation}
is to be determined. It can be derived that $\psi_c = 0$ with the $\delta$-function distribution, i.e.\ $S=1-H\left(r-r_c\right)-r_c \delta\left(r-r_c\right)/2$, where $H$ is the Heaviside step function, and $\delta$ is the Dirac delta function. But in this work, $S$ is finite everywhere, so that $\psi_c$ is non-zero, and $\partial \psi/\partial r$ is continuous everywhere because of Eq.~(\ref{eq:psi_S}), which means
\begin{equation}
    \left.\frac{\partial}{\partial r}\psi\right|_{r=r_c} = - \frac{r_c}{2}. \label{eq:left_bound}
\end{equation}

For a plasma electron which is at rest before the drive beam arrives, there is a constant of motion~\cite{PMoraPOP1997}
\begin{equation}
    \gamma - v_z \gamma - \psi = 1, \label{eq:const_motion}
\end{equation}
where $\gamma=1/\sqrt{1-v_z^2-v_r^2}$ is the Lorentz factor, $v_z$ and $v_r$ are the longitudinal and radial components of the velocity of the electron, respectively. This equation can be solved under the adiabatic assumption, which means $v_r\approx 0$, as
\begin{equation}
    v_z = \frac{2}{1+\left(1+\psi\right)^2} -1. \label{eq:vz}
\end{equation}

Based on the adiabatic assumption, the transverse electromagnetic field can be written using the Gauss' Law and Stokes' Theorem
\begin{eqnarray}
    E_r &=& \frac{r}{2} - \frac{\Lambda}{r} - \frac{1}{r}\int_0^r n_e\left(r'\right) r' dr',\\
    B_\theta &=& - \frac{\Lambda}{r} - \frac{1}{r}\int_0^r n_e\left(r'\right) v_z\left(r'\right) r' dr'.
\end{eqnarray}
The transverse force $F_r = -E_r + v_z B_\theta$ should be balanced for $r>r_c$, which means
\begin{equation}
\begin{aligned}
    0 = rF_r =& -\frac{r^2}{2} + \left(1-v_z\right)\Lambda + \int_0^r n_e\left(r'\right) r' dr'\\
    &- v_z\int_0^r n_e\left(r'\right) v_z\left(r'\right) r' dr'.
\end{aligned} \label{eq:balance}
\end{equation}
The channel radius is determined by letting $r=r_c$ in Eq.~(\ref{eq:balance}), as
\begin{equation}
    r_c = \sqrt{2\left(1-v_{zc}\right)\Lambda} = 2\sqrt{\frac{\Lambda}{1+\frac{1}{\left(1+\psi_c\right)^2}}}, \label{eq:rc}
\end{equation}
where
\begin{equation}
    v_{zc} = \left.v_z\right|_{r=r_c} = \frac{2}{1+\left(1+\psi_c\right)^2} -1. \label{eq:vzc}
\end{equation}
We see that $r_c$ reduces to $r_n$ if $\psi_c=0$, and to $r_{\delta0}$ if $\psi_c\rightarrow\infty$. Then, we know $r_n < r_c <r_{\delta0}$ if $\psi_c$ is non-zero and finite, because $r_c$ monotonically increases with $\psi_c$.

The sheath equations defined by Eqs.~(\ref{eq:psi_S}), (\ref{eq:vz}) and (\ref{eq:balance}) are solvable using the left boundary condition Eq.~(\ref{eq:left_bound}) and the right boundary condition
\begin{equation}
    \lim_{r\to\infty}\psi = 0, \label{eq:right_bound}
\end{equation}
which is a natural requirement that the plasma disturbance is local.

\begin{figure}
    \centering
    \begin{overpic}[width=0.483\textwidth]{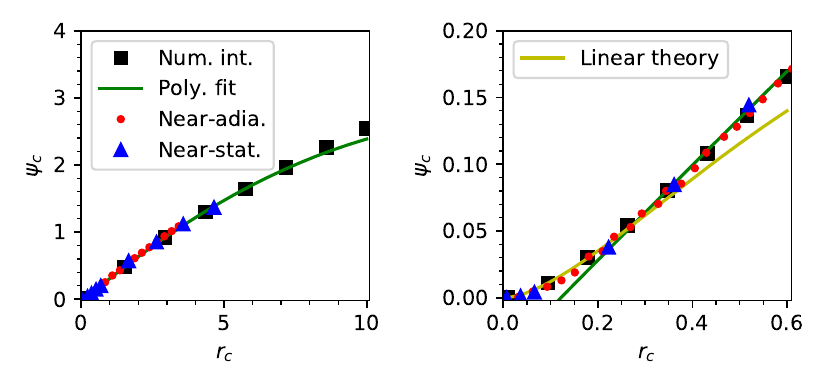}
       \put(39,13){\sffamily (a)}
       \put(90,13){\sffamily (b)}
    \end{overpic}
    \caption{\label{fig:psic_rc} Comparison of $\psi_c$ vs $r_c$ obtained by the numerical integral of Eqs.~(\ref{eq:psi_S}), (\ref{eq:vz}) and (\ref{eq:balance}) using the boundary conditions Eqs.~(\ref{eq:left_bound}) and (\ref{eq:right_bound}) (black squares), by the polynomial fit Eq.~(\ref{eq:psic0}) (green line), by the linear theory Eq.~(\ref{eq:psic1}) (yellow line), by PIC simulations in near-adiabatic cases (red dots) and in near-stationary cases (blue triangles). Both the large (a) and small (b) scales of $r_c$ are plotted, in order to show the different asymptotic behavior of $\psi_c$ at the two ends.}
\end{figure}

We use the shooting method to numerically solve the equations. In other words, the numerical integral is performed from $r=r_c$ to a sufficient large value of $r$, while $\psi_c$ is predicted and corrected in iterations, so that Eq.~(\ref{eq:right_bound}) is satisfied to a certain accuracy. Practically, there is a critical value of $\psi_c$ for each $r_c$, which prevents divergence at large $r$. The Python script for the numerical solution is attached as a Supplementary Material, and is maintained on GitHub~\cite{MZengAdiabaticSheathsolver2024}. The numerical solution of $\psi_c$ with different $r_c$ is shown in Fig.~\ref{fig:psic_rc}. The polynomial fit,
\begin{equation}
    \psi_c \approx -0.012 r_c^2 + 0.363 r_c - 0.044 , \label{eq:psic0}
\end{equation}
is a good estimate for $r_c \lesssim 8$ as shown in Fig.~\ref{fig:psic_rc}(a).

After $\psi_c$ is obtained, the electron density at the sheath boundary can be predicted. By taking derivative of Eq.~(\ref{eq:balance}), we obtain
\begin{equation}
    n_e r \left(1-v_z^2\right) = r + \Lambda\frac{\partial v_z}{\partial r} + \frac{\partial v_z}{\partial r}\int_0^r n_e\left(r'\right) v_z\left(r'\right) r' dr'. \label{eq:balance_derivative}
\end{equation}
At $r=r_c$, the integral is zero, the derivative of $v_z$ can be determined by Eqs.~(\ref{eq:left_bound}) and (\ref{eq:vz}), and $\Lambda$ can be written as the function of $r_c$ and $\psi_c$ according to Eq.~(\ref{eq:rc}). As a result,
\begin{equation}
    \left.n_e\right|_{r=r_c} = \frac{r_c^2}{8}\left(\frac{1}{\psi_*}+\frac{1}{\psi_*^3}\right) + \frac{1}{4}\left(\psi_* + \frac{1}{\psi_*}\right)^2,
\end{equation}
where $\psi_* = 1+\psi_c$.

Next, we study the behavior of $\psi$ and $\psi_c$ in the limit $\psi_c\ll 1$. This also means $r_c\ll 1$, $\Lambda\ll 1$, and $\psi\ll 1$. By expanding Eqs.~(\ref{eq:vz}) and (\ref{eq:balance_derivative}) and keeping only the 1st order terms of $\psi$, we have $v_z\approx -\psi$ and $n_e \approx 1$. Thus, the linear form of Eq.~(\ref{eq:psi_S}) is obtained
\begin{equation}
    \frac{\partial^2}{\partial r^2} \psi + \frac{1}{r} \frac{\partial}{\partial r} \psi - \psi =0.
\end{equation}
The solution satisfying the right boundary condition Eq.~(\ref{eq:right_bound}) is $\psi = R\left(r_c\right) K_0\left(r\right)$, where $R\left(r_c\right)$ is a constant of $r$ to be determined, and $K_0\left(r\right)$ is the modified Bessel function of the second kind of order zero~\cite{AJeffrey2008}. By also considering the left boundary condition Eq.~(\ref{eq:left_bound}), we may get the expression of $R\left(r_c\right)$ to the lowest order of $r_c$, and find
\begin{equation}
    \left.\psi\right|_{r>r_c} = \dfrac{r_c^2}{2} K_0\left(r\right).\label{eq:psi_linear}
\end{equation}
Thus,
\begin{equation}
    \psi_c = \dfrac{r_c^2}{2} K_0\left(r_c\right), \label{eq:psic1}
\end{equation}
or $\psi_c \approx -\frac{r_c^2}{2} \left( \ln \frac{r_c}{2} + 0.577...\right)$, where the second summand is the Euler-Mascheroni constant. We can see in Fig.~\ref{fig:psic_rc}(b) that Eq.~(\ref{eq:psic1}) is satisfactory with the numerical results for $r_c \lesssim 0.28$, while Eq.~(\ref{eq:psic0}) is better for $r_c \gtrsim 0.28$.

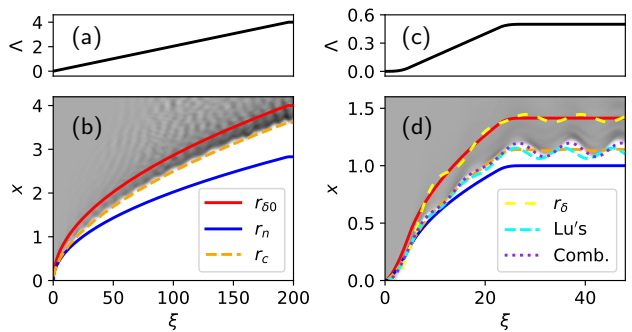
\begin{figure}
    \centering
    \begin{overpic}[width=0.48\textwidth]{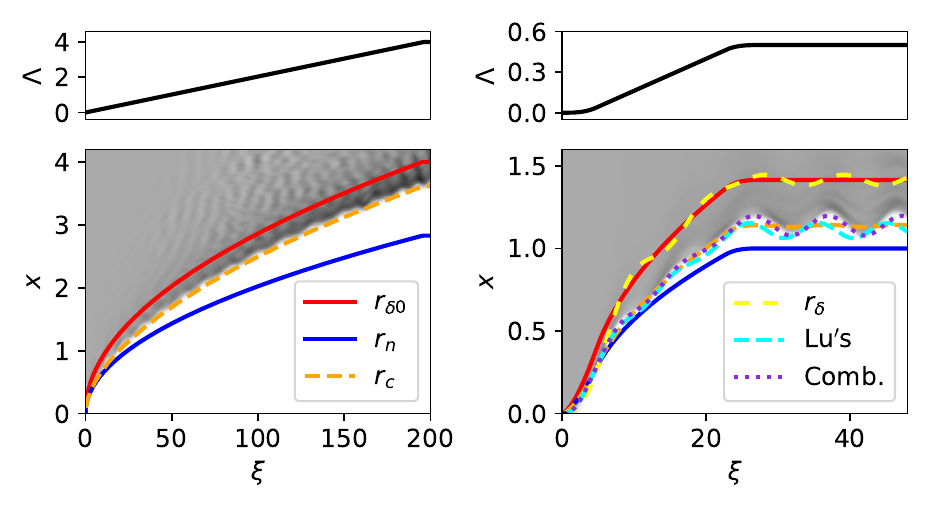}
       \put(12,47){\sffamily (a)}
       \put(12,33){\sffamily (b)}
       \put(63,47){\sffamily (c)}
       \put(63,33){\sffamily (d)}
    \end{overpic}
    \caption{\label{fig:rc_vs_xi} The distribution of $\Lambda$ and slice plots of plasma electron density at $y=0$ (gray) in simulations for (a) and (b) a near-adiabatic case, and (c) and (d) a near-stationary case. The numerical solution of the balancing radius based on the $\delta$-sheath theory $r_\delta$ (yellow dashed line) and its balance $r_{\delta0}$ (red line), the neutralization radius $r_n$ (blue line), the solution based on our theory $r_c$ (orange dashed line), the numerical solution based on Lu's model with $\beta = 1/r_c+0.1$ (cyan dashed line) and the combined model with $\beta$ in Lu's model replaced by Eq.~(\ref{eq:beta_psic}) (violet dotted line) are plotted for the comparison.}
\end{figure}

Once $\psi_c$ is determined, the one-to-one mapping between $\Lambda$ and $r_c$ can be obtained according to Eq.~(\ref{eq:rc}). For comparison, PIC simulations have been performed using the quasi-static code \texttt{QuickPIC}~\cite{CKHuangJCP2006,WMAnJOCP2013}. In the simulations, we consider a 5 GeV drive electron beam, transversely positioned at $x=y=0$ (i.e. the center of the simulation domain). It has a sufficiently small transverse size, while its longitudinal size is the same as the entire simulation domain. Two sets of simulations with different drive beam current profiles have been conducted. One is called the near-adiabatic profile, where $\Lambda$ increases linearly with a sufficiently small slope. The other is called the near-stationary profile, where $\Lambda$ increases from 0 to a plateau, and the increasing profile and length are designed so that the oscillation of the channel radius in the plateau is not severe.

In the near-adiabatic simulations, we have chosen the maximum $\Lambda_{\max}= 4$ for the large $r_c$-scale shown in Fig.~\ref{fig:psic_rc}(a) and in Fig.~\ref{fig:rc_vs_xi}(a), and $\Lambda_{\max}= 0.4$ for the small $r_c$-scale shown in Fig.~\ref{fig:psic_rc}(b). The simulation domain has a size of 200 in $\xi$-direction, which is the propagation direction, and 40 (10) in $x$- and $y$-direction for the $\Lambda_{\max}=4$ (0.4) case. The numbers of cells is chosen as 2048, 1024 and 1024, respectively. The near-adiabatic simulation results for $\psi_c$ vs.\ $r_c$ are shown in Fig.~\ref{fig:psic_rc}, which are in good agreements with our model.

In the near-stationary simulations, $\Lambda$ increases from 0 to the plateau value $\Lambda_0$ within $0\leq \xi \leq 25$, and keeps the plateau value in the remaining of the simulation domain, as shown in Fig.~\ref{fig:rc_vs_xi}(c). Cubic splines have been designed to smooth the transitions near $\xi = 0$ and 25. Here, the simulation domain has a longitudinal size of 48, while the transverse size is adjusted in accordance with $\Lambda_0$, such that $\psi\rightarrow 0$ at the simulation boundary is guaranteed ($\partial \psi/\partial r \approx 0$ at the boundary), and the transverse resolution is adequate to resolve the sheath. Again, the number of cells is chosen to be $2048 \times 1024 \times 1024$. $\Lambda_0$ has been scanned from $2.5 \times 10^{-5}$ to 6.25 in the simulations. The results of $\psi_c$ vs.\ $r_c$ are shown in Fig.~\ref{fig:psic_rc}, which are also in good agreements with our model.

Plasma density slices at $y=0$ for two of
the above mentioned simulations are shown in Fig.~\ref{fig:rc_vs_xi}(b) and \ref{fig:rc_vs_xi}(d). We compare the simulations with $r_{\delta0}$ obtained by Eq.~(\ref{eq:r_delta0}), $r_n$ obtained by Eq.~(\ref{eq:r_n}) and $r_c$ obtained by Eq.~(\ref{eq:rc}). We see that the channel radius in simulations best matches with $r_c$, while $r_n$ is an underestimate and $r_{\delta0}$ is an overestimate.

Furthermore, the channel radius slightly oscillates around $r_c$ in the near-stationary case because of the imperfect balancing, as shown in Fig.~\ref{fig:rc_vs_xi}(d). This oscillation behavior can be numerically solved using Lu's model~\cite{WLuPOP2006, WLuPRL2006}, which is similar to Eq.~(\ref{eq:r_delta}), but the parameters are $A=1+\left[1/4 + \beta/2 + \left(r_c/8\right) \left(d\beta/dr_c\right)\right] r_c^2$, $B=1/2 + \left(3/4\right) \beta + \left(3 r_c/4\right) \left(d\beta/dr_c\right) + \left(r_c^2/8\right) \left(d^2 \beta/dr_c^2\right)$ and $C=\left[1 + 1/\left(1 + \beta r_c^2/4\right)^2\right]/4$. The variable $\beta$ needs prior knowledge to the sheath thickness, and was recommended to be $\beta = 1/r_c+0.1$. Although Lu's model is in good agreement with the simulation in the short drive cases (drive beam is shorter than or comparable to the plasma wavelength), it does not predict the oscillation of $r_c$ very well in the long drive cases as shown in Fig.~\ref{fig:rc_vs_xi}(d).

To improve Lu's model, we replace $\beta$ by
\begin{equation}\label{eq:beta_psic}
    \beta = \frac{4\psi_c}{r_c^2},
\end{equation}
where $\psi_c$ is expressed in a piecewise manner by Eq.~(\ref{eq:psic1}) for $r_c<0.28$ and by Eq.~(\ref{eq:psic0}) for $r_c>0.28$, and call it the combined model. The combined model has been applied to the near-stationary cases, and its numerical solutions have the best agreement with the simulations compared with other models, with one example shown as the violet dotted line in Fig.~\ref{fig:rc_vs_xi}(d).

In conclusion, we have introduced a self-consistent model for the sheath structure of the blowout plasma channel driven by an electron beam based on the adiabatic assumption. The model consists of the equation for the pseudo-potential Eq.~(\ref{eq:psi_S}), the longitudinal velocity of the sheath electrons Eq.~(\ref{eq:vz}), and the transverse force balancing condition Eq.~(\ref{eq:balance}), which is solvable with the boundary conditions Eqs.~(\ref{eq:left_bound}) and (\ref{eq:right_bound}). An analytical solution is obtained in the small-blowout limit as Eq.~(\ref{eq:psi_linear}), and a general solution is retrieved numerically by the shooting method. The PIC simulations show that the radius obtained by Eq.~(\ref{eq:rc}) is a better estimate of the channel balancing radius than that from previous neutralization or $\delta$-sheath models.

It should be noted that the accuracy of our model is limited in short drive beam cases. Nevertheless, our model is very suitable for long drive beams. Especially, by combining our model with Lu's model, we have obtained the most correct description of the oscillation of the channel radius. The better understanding of the pseudo-potential at the sheath boundary and inside the blowout for long drive cases is crucial for high transformer ratio and future high-energy PWFA studies~\cite{JGaoRDTM2024,LuPAC2010,LoischGPRL2018}.

\begin{acknowledgments}
This work is supported by the Strategic Priority Research Program of the Chinese Academy of Sciences (Grant No.~XDB0530000), the National Natural Science Foundation of China (Grant No.~12475159), and the BMBF project 05P24PF2 (Germany).
\end{acknowledgments}
\bibliography{main}
\end{document}